\documentclass[fleqn,usenatbib]{mnras}
\usepackage{newtxtext,newtxmath}
\usepackage[T1]{fontenc}
\usepackage{graphicx}
\usepackage{amsmath}
\usepackage{siunitx}
\title[Self-calibration of weak lensing cosmic shear biases]{Self-calibration of weak lensing cosmic shear biases}

\author[Giuseppe Congedo \& Andy Taylor]{
Giuseppe Congedo,$^{1}$\thanks{E-mail: \href{mailto:giuseppe.congedo@ed.ac.uk}{giuseppe.congedo@ed.ac.uk}}
Andy Taylor,$^{1}$
\\
$^{1}$Institute for Astronomy, University of Edinburgh, Royal Observatory, Blackford Hill, Edinburgh EH9 3HJ, UK \\
}

\date{Accepted XXX. Received YYY; in original form ZZZ}

\pubyear{\the\year{}}

\defcitealias{congedo2024}{C24}

\begin{document}
\label{firstpage}
\pagerange{\pageref{firstpage}--\pageref{lastpage}}
\maketitle

\begin{abstract}
In order to reach the required performance of Stage-III and IV weak lensing surveys, cosmic shear measurements have to rely on external simulations to calibrate residual biases.
Over the years, several techniques have been developed to mitigate the impact of residual biases prior to calibration,
including the inference of shear responses on images to correct multiplicative biases, and the empirical correction of additive biases.
We introduce a novel methodology that generalises upon the state-of-the-art approaches by inferring multiplicative and additive biases jointly from parameterised distributions of measured ellipticities, crucially without relying on external simulations and independently from cosmology.
Shear biases are marginalised over the unknown hyper-parameters in the modelling, hence mitigating the impact of degeneracies.
We apply the technique to a representative problem and show the performance of the estimation, even in the presence of noise.
The method has a high potential for applicability to the calibration of weak lensing cosmic shear in current and future lensing surveys.
\end{abstract}


%%%%%%%%%%%%%%%%%%%%%%%%%%%%%%%%%%%%%%%%%%%%%%%%%%

\section{Introduction}

With the advent of Stage-IV galaxy surveys represented by the three main wide-field telescopes [\textit{Euclid}, \citet{mellier2024}; \textit{Rubin}, \citet{ivezic2019}; and \textit{Roman}, \citet{spergel2015}], cosmological measurements will soon enter the regime of percent-level precision.
One of the main cosmological probes is weak lensing cosmic shear, the distortion of images of background galaxies due to the large-scale structure in the foreground \citep{schneider2005,kilbinger2015}.
Although weak lensing is a promising technique for constraining the cosmological parameters of dark matter, dark energy and modified gravity, Stage-III surveys have had to correct systematic errors with simulations in order to reach the desired accuracy of cosmological measurements \citep{fenech-conti2017,mandelbaum2018,kannawadi2019,maccrann2022,li2022,li2023a,li2023b}.

One of the key sources of systematic errors is the modelling of the telescope's point-spread function (PSF), since residual errors propagate through the inferred cosmic shear.
In fact, PSF ellipticity errors induce additive biases in galaxy measurements, whereas PSF size errors induce multiplicative biases \citep{paulin-henriksson2008}.
Moreover, detector effects also introduce additional additive biases, such as for the case of charge transfer inefficiency \citep{rhodes2010}, but sometimes also multiplicative errors, such as for the case of the bright-fatter effect \citep{antilogus2014}.
Even in the idealistic situation where the PSF might be perfectly known, there are always multiplicative biases introduced by the measurement methodology (\citealt{miller2013}; \citealt{congedo2024}, hereafter \citetalias{congedo2024}) or due to the non-uniform distribution of galaxies (\citealt{martinet2019,maccrann2022}; \citetalias{congedo2024}).

Over the years, much effort has been put initially towards understanding the source of fundamental biases due to the measurement process \citep{melchior2012,refregier2012,hall2017} and later towards trying to correct measurement and selection biases directly in the measurement process itself \citep{bernstein2014,sheldon2014,bernstein2016,huff2017,sheldon2017,sheldon2020,hoekstra2021a,hoekstra2021b}.
However, with much stringent lensing accuracy required for a Stage-IV survey like \textit{Euclid} \citep{massey2013,cropper2013}, residual biases still have to be corrected for via simulations, which raise questions about the sensitivity of the calibration on the simulation setup or the data volume required for the correction (\citealt{hoekstra2017,pujol2019,jansen2024}; \citetalias{congedo2024}).

Residual additive shear biases are known to be more forgiving in that they can be empirically corrected directly on the data, which allow us to capture their spatial dependence \citep{vanuitert2016,hildebrandt2020,kitching2021}.
In fact, this methodology is general enough to be easily applied to any broad sample, including tomographic, field-of-view, angular, or morphological bins.
However, multiplicative biases are notoriously much harder to correct and recent developments focussed primarily on estimating the sensitivity of the shear measurements and selections to an input shear, with a procedure called \textsc{Metacalibration}/\textsc{Metadetection} yielding shear responses as direct proxies for multiplicative biases \citep{huff2017,sheldon2017,sheldon2020,sheldon2023}.
However, there is sensible expectation that even those methodologies may not be fully immune to residual biases, especially when applied to Stage-IV surveys where the requirements and precision are much more stringent than Stage III, and therefore simulations are still required to correct for residual biases.

In this paper, we propose a novel method to estimate residual additive and multiplicative biases in a unified, general approach.
Our methodology infers multiplicative and additives biases directly from the observed distribution of ellipticity and without relying on assumptions about morphologies or cosmology.
Our paper is structured as follows: Sec.\;\ref{sec:distributions} presents the general properties of the transformation of lensed ellipticity distributions under the effect of biases; Sec.\;\ref{sec:inference} introduces the Bayesian inference of multiplicative and additive biases from empirical distributions of observed ellipticity while marginalising over nuisance parameters; Sec.\;\ref{sec:analysis} verifies the procedure on representative data and shows the performance of our estimator, while also discussing potential limitations; Sec.\;\ref{sec:conclusions} draws the main conclusions of our work.

\section{Biased ellipticity distributions}\label{sec:distributions}

We wish to understand the properties of the transformation between the distribution of unbiased lensed ellipticities and that of measured ellipticities under the effect of shear biases and noise.
We define biases as constant `effective' quantities for a broad sample selection.
In reality, lensing biases will respond differently depending on, e.g., galaxy morphology, brightness, or ellipticity.
In our modelling, we are only interested in the sensitivity of the distribution \textit{as a whole} to biases and noise for a broad sample of galaxies responding differently to those effects. 
In this context, the \textit{unbiased lensed} galaxy ellipticity, $\epsilon$, is a spin-2 complex number representing the lensed ellipticity of a galaxy before the measurement \citep{seitz1997},
\begin{equation}\label{eq:shear}
\epsilon=\frac{\epsilon_\text{s}+g}{1+\epsilon_\text{s}\,g^*}~,
\end{equation}
where $\epsilon_\text{s}$ is the intrinsic ellipticity of the source galaxy, $g$ is the reduced shear, $g=\langle\epsilon\rangle$, and $\epsilon\approx\epsilon_\text{s}+g$ in the weak-lensing regime \citep{schneider2005,kilbinger2015}.
After the measurement, the \textit{measured} or \textit{observed} galaxy ellipticity, $\hat{\epsilon}$, will include statistical noise and bias, meaning that it is the result of a noisy, biased measurement process \textit{after} the effect of cosmic shear.
With both $\epsilon_\text{s}$ and $g$ being zero-mean random variables, $\epsilon$ follows a symmetric distribution centred at zero with $|\epsilon|<1$, except in the case of intrinsic alignments \citep{joachimi2015}, which should have negligible effect on the mean of a broad sample like a tomographic bin.

The most general bias transformation linking measured shear, $\hat{g}$, and true shear, $g$, is given in terms of a spin-0 or 4 multiplicative biases, $m$, and a spin-2 additive bias, $c$ (\citealt{kitching2021}; \citetalias{congedo2024}).
For simplicity, in this paper, we assume a simpler 1D linear model with scalar biases \citep{guzik2005,huterer2006,heymans2006},
\begin{equation}\label{eq:shear_bias}
\hat{g}=(1+m)\,g+c+g_\text{n}~,
\end{equation}
where $m$ and $c$ are effective biases, i.e., real scalars defined over a large sample of galaxies, including measurement and selection effects, and $g_\text{n}$ is zero-mean statistical noise for the whole selection.
Because the ellipticity is a point estimator for shear and $\hat{g}=\langle\hat{\epsilon}\rangle$ and $g=\langle\epsilon\rangle$ for large numbers of galaxies, a linear model of how the ellipticities transform under effective shear biases for the entire sample is as follows,
\begin{equation}\label{eq:ellipticity_bias}
\hat{\epsilon}=(1+m)\,\epsilon+c+\epsilon_\text{n}~,
\end{equation}
where $m$ and $c$ are effective constants and $\epsilon_\text{n}$ is zero-mean statistical noise.
The actual relation between measured and unbiased ellipticity for individual galaxies will not, in general, follow Eq.\;\eqref{eq:ellipticity_bias} due to the non-linear response of the ellipticity to a number of effects.
However, for broad samples, the many effects happening in real data will compound or cancel out, leading to total multiplicative biases at the percent level.
In this limit of small cosmic shear biases and large enough signal-to-noise, the bias relation can be safely assumed to be linear. 
%We will see that edge effects, which are expected to induce non-linear effects, are down-weighted by our analysis.

\begin{figure}
\centering
\includegraphics[width=\columnwidth]{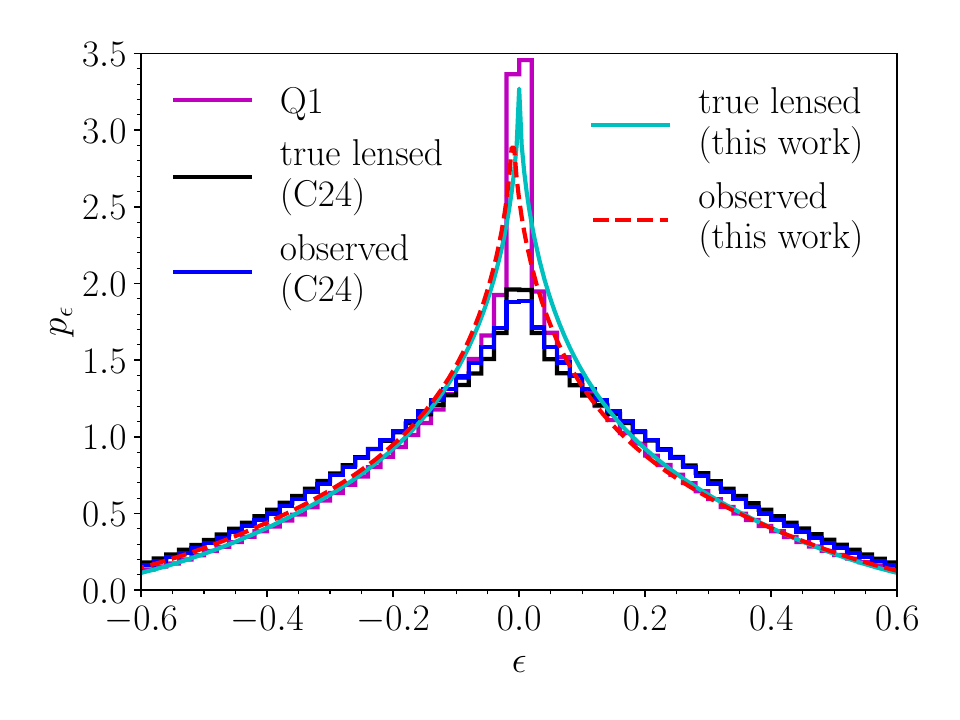}
\vspace{-20pt}
\caption{Distribution of ellipticity, observed or modelled.
`Q1': distribution of observed ellipticity from the \textit{Euclid} Quick Release 1 \citep{congedo2026}.
`C24': unbiased lensed and observed ellipticity showing the negligible impact of measurement noise and biases in our pre-launch simulations \citepalias{congedo2024}.
Finally, `this work': unbiased lensed and observed ellipticity after the effect of large shear biases of $m=0.05$ and $c=-0.01$.}
\label{fig:distributions}
\end{figure}

According to Eq.\;\eqref{eq:ellipticity_bias}, the observed 1D-probability distribution of measured ellipticity, $p_{\hat{\epsilon}}(\hat{\epsilon})$, is related to the unbiased lensed distribution, $p_\epsilon(\epsilon)$, via the conservation of total probability,
\begin{equation}\label{eq:cons_prob}
p_{\hat{\epsilon}}\left(\hat{\epsilon}|m,c\right) = \frac{1}{|1+m|}p_\epsilon\left(\frac{\hat{\epsilon}-c}{1+m}\right)*p_{\epsilon_\text{n}}(\epsilon_\text{n})~,
\end{equation}
where $p_{\epsilon_\text{n}}(\epsilon_\text{n})$ is the distribution of ellipticity noise.
Figure\;\ref{fig:distributions} illustrates how flexible our model is in reproducing trends in both real data and simulations.
For real data, it includes an example from the \textit{Euclid} Quick Release 1 \citep{congedo2026}.
Also, it shows the negligible impact of noise and biases in our pre-launch simulations \citepalias{congedo2024}, which used the Flagship simulations employing $N$-body simulations of dark matter halos and galaxies \citep{castander2025}, and the effect that biases $m=0.05$ and $c=-0.01$ can have on the shape of the observed distribution, whose model is that of Eq.\;\eqref{eq:model}. 
In general, $c$ shifts the mean of the distribution and $m$ rescales the distribution by expansion or contraction, depending on its sign.
Additionally, the noise broadens the distribution via a convolutive term.
However, the convolution with a random process and the affine/linear operation of biases are fundamentally different in their nature:
the former is not reversible while the latter is fully deterministic and, therefore, reversible.
Hence, accurate forward modelling could allow breaking the degeneracy and distinguishing between the two.
It is worth noting that the observed distribution in Eq.\;\eqref{eq:cons_prob}, while exact and a direct consequence of Eq.\;\eqref{eq:ellipticity_bias}, holds true only for the population as a whole.
It does not apply to individual galaxies that will be biased very differently from the average galaxy in the sample.
For example, galaxies falling in the tail of the distribution will naturally have large ellipticities and will suffer from non-linearities.
Luckily, these effects are automatically down-weighted by our analysis presented in the next section.

The ellipticity distribution is highly non-Gaussian. Empirical distributions show a sharp, cusp-like peak at zero and negative excess kurtosis.
This suggests that information on $m$ and $c$ can potentially be extracted at all higher-order moments.
It is worth studying what happens to the moments, both central and raw, under the effect of an affine transformation, at least in the limit of negligible noise.
The generalisation with random noise proceeds in a similar way, but the equations will be more complicated, showing potential correlations between biases and the noise moments.
However, we will see that our forward modelling approach overcomes the issues.
The $k$-th central moment of the measured ellipticity distribution is
\begin{equation}
\mu^0_k[p_{\hat{\epsilon}}]=(1+m)^k\mu^0_k[p_\epsilon]~,
\end{equation}
where the $k$-th central moment of the unbiased lensed ellipticity distribution is defined by
\begin{equation}
\mu^0_k[p_\epsilon]=\langle(\epsilon-\langle\epsilon\rangle)^k\rangle~.
\end{equation}
As is clear, central moments, by definition, are insensitive to $c$, but depend on powers of $1+m$.
For example, measuring the variance ($k=2$) could, in principle, be sufficient to estimate $m$ if accurate knowledge of the unbiased ellipticity distribution were available.
However, skewness and excess kurtosis,
\begin{align}
\gamma_1&=\frac{\mu^0_3}{(\mu^0_2)^{3/2}}~, \\
\gamma_2&=\frac{\mu^0_4}{(\mu^0_2)^{2}}-3~,
\end{align}
are insensitive to $m$ because the same $1+m$ cancels out at the numerator and denominator in their definitions.
Let us focus on the raw moments of the measured ellipticity distribution defined as follows
\begin{equation}
\mu_k[p_{\hat{\epsilon}}]=\sum_{j=0}^k\binom{k}{j}(1+m)^j c^{k-j}\mu_j[p_\epsilon]~,
\end{equation}
where the $k$-th raw moment of the unbiased lensed ellipticity distribution is defined by
\begin{equation}
\mu_k[p_\epsilon]=\langle\epsilon^k\rangle~.
\end{equation}
Unlike central moments, raw moments are sensitive to a combination of $m$, $c$, and noise.
In the case of a centred Gaussian distribution, all odd moments are zero and all cumulants beyond the mean and standard deviation are also zero.
Therefore, a Gaussian distribution is fully described in terms of its mean and standard deviation and contains very limited information on $m$ and $c$ compared to a more realistic distribution.

The $k$-th raw moments are also defined in terms of the $k$-th Taylor coefficients of the moment-generating function (MGF).
Therefore, it makes sense to derive the relation between the MGFs for the unbiased lensed and measured ellipticity distributions,
\begin{equation}\label{eq:mgf}
M_{\hat{\epsilon}}(t)=e^{t\,c}\langle e^{t\,\epsilon_\text{n}}\rangle M_{\epsilon}^{1+m}(t)~,
\end{equation}
where the MGF of the unbiased lensed ellipticity distribution is given by
\begin{equation}
M_\epsilon(t)=\langle e^{t\epsilon}\rangle~,
\end{equation}
or, equivalently, defined as the double-sided Laplace transform of $p_\epsilon$.
Also, the noise term, $\epsilon_\text{n}$, is assumed to be uncorrelated from $\epsilon$.
As is clear from Eq.\;\eqref{eq:mgf}, the MGF of the measured ellipticity distribution reduces to the MGF of the unbiased lensed ellipticity distribution in the limit of negligible noise and biases.
In the Fourier domain, the combined effect of $c$ and noise is to induce a modulation, while $m$ changes the compactness of the distribution.
The Taylor expansion of Eq.\;\eqref{eq:mgf} yields the exact moments of the observed ellipticity distribution and their dependence on $m$, $c$, and noise to any order, which we now aim to exploit.

\section{Inferring biases from data}\label{sec:inference}

We know that $m$ is potentially degenerate with the dispersion of the unbiased lensed ellipticity distribution.
If the unbiased dispersion were accurately known, then the degeneracy between $m$ and dispersion could be broken.
This observation suggests an effective approach to the problem of inferring biases from the data.
Ellipticities (whether unbiased, measured, or lensed) are always bound to $|\epsilon|<1$ as not to invalidate the property of an $\epsilon$-ellipticity and its transformation under shear.\footnote{The $\epsilon$-ellipticity of a galaxy is defined as $\epsilon=(a-b)/(a+b)\exp{2i\phi}$, where $a$, $b$, and $\phi$ are semi-major axis, semi-minor axis, and rotation angle of the galaxy.
This definition, under the shear transformation of Eq.\;\eqref{eq:shear}, preserves the support of the distribution with $|\epsilon|<1$.}
Also, observed ellipticity distributions will be symmetric, although around a non-zero mean, because skewness is invariant under the bias transformation.
Additionally, the response of the observed distribution on biases will generally be different in regions around zero or large ellipticities, but it is the cumulative effect on the ellipticity distribution as a whole that matters for shear calibration, not how galaxy measurements respond individually to morphologies or systematics.
All these statistical properties, and in particular the sensitivity of the measured ellipticity distribution to biases to any higher-order moments, help to lift the degeneracy in $m$.

The problem of estimating shear biases from data can be fully framed in Bayesian terms by forward modelling the effect that biases can have on the distribution as a whole without limiting ourselves to just low-order moments.
Let $\hat{p}_{\hat{\epsilon},i}$ be the empirical probability distribution function of the measured ellipticities for a relatively large, representative sample, and $\hat{\epsilon}_i$ the observed ellipticity of the $i$-th bin.\footnote{The relation between the empirical distribution and the standard histogram count is given by the number of samples multiplied by the bin width, such that the empirical distribution integrates to one.} 
This empirical distribution is the data we wish to fit with a model for the unbiased lensed distribution, $p_\epsilon(\epsilon|\theta)$, where $\theta$ is the vector of hyper-parameters.
We must also include the bias transformation in the modelling, to transform from the unbiased distribution to the observed, biased distribution.
The probability of observing the `data' distribution, $\hat{p}_{\hat{\epsilon},i}$, given the `model', $p_{\hat{\epsilon}}\left(\hat{\epsilon_i}|m,c,\theta\right)$ parameterised according to Eq.\;\eqref{eq:cons_prob}, is denoted with $\pi(\hat{p}_{\hat{\epsilon}}|m,c,\theta)$ and is a likelihood of the model parameters, $m$, $c$, and $\theta$.
Since bin counts follow a Poisson distribution, this likelihood is the product of many individual Poisson distributions,
\begin{equation}\label{eq:likelihood}
\ln\pi(\hat{p}_{\hat{\epsilon}}|m,c,\theta)=N\,\Delta\sum_i \hat{p}_{\hat{\epsilon},i}\ln p_{\hat{\epsilon}}(\hat{\epsilon_i}|m,c,\theta)+\text{const}~,
\end{equation}
where $N$ is the number of samples and $\Delta$ is the bin width.
The likelihood includes the additional constraint on the model, $\Delta\sum_i p_{\hat{\epsilon}}(\hat{\epsilon_i}|m,c,\theta)=1$
and the term $\ln(N\,\Delta\,p_{\hat{\epsilon},i}!)$, both absorbed by the constant.
Thanks to the logarithm, the likelihood naturally down-weights the points with low probability or low counts.
According to the central limit theorem, the likelihood converges to a Gaussian very rapidly, but in general the Poisson assumption remains more accurate, particularly for small number of samples.

By the Bayes theorem, the joint posterior distribution of shear biases and distribution hyper-parameters, given the empirical distribution, is
\begin{equation}\label{eq:posterior}
\pi(m,c,\theta|\hat{p}_{\hat{\epsilon}}) = \frac{\pi(\hat{p}_{\hat{\epsilon}}|m,c,\theta)\,\pi(m,c,\theta)}{\pi(\hat{p}_{\hat{\epsilon}})}~,
\end{equation}
where $\pi(m,c,\theta)$ is the joint prior distribution of shear biases and hyper-parameters, and  $\pi(\hat{p}_{\hat{\epsilon}})$ is the marginal likelihood, i.e. the integral of Eq.\;\eqref{eq:likelihood} over $m$, $c$, and $\theta$.
The final posterior distribution of shear biases after marginalising over the unknown hyper-parameters is
\begin{equation}
\pi(m,c|\hat{p}_{\hat{\epsilon}}) = \frac{1}{\pi(\hat{p}_{\hat{\epsilon}})}\int\pi(\hat{p}_{\hat{\epsilon}}|m,c,\theta)\,\pi(m,c,\theta)\,\text{d}\theta~.
\end{equation}
The problem is now equivalent to constraining $m$ and $c$ statistically from an observed ellipticity distribution given a forward model for the unbiased lensed ellipticity distribution.

\section{Analysis}\label{sec:analysis}

We adopted a relatively flexible model for the lensed 1D-ellipticity distribution,
which is based on our pre-launch simulations \citepalias{congedo2024} and more recent results from the analysis of the Q1 data \citep{congedo2026}, as shown in Fig.\;\ref{fig:distributions}.
The functional form of the model is as follows,
\begin{equation}\label{eq:model}
p_\epsilon(\epsilon|\sigma,\beta,\kappa) \propto
\exp\left[
-\left(\frac{|\epsilon|}{\sigma}\right)^\beta
-\left(\frac{|\epsilon|}{\alpha\,\sigma}\right)^{2\,\gamma}\right]
\left[1 - \left(\frac{|\epsilon|}{\epsilon_{\max}}\right)^\kappa\right]~.
\end{equation}
The exponential is a bell-like function with a dispersion ($\sigma$), exponential decay ($\beta$), and quadratic-exponential decay ($\gamma$).
The additional tapering function with a decay coefficient ($\kappa$) ensures that the distribution is zero at a fixed $\epsilon_{\max}=0.99$, which should coincide with the upper bound of the actual measurements \citepalias{congedo2024}.
The vector of hyper-parameters that we wish to marginalise over in the estimation of bias parameters is then $\theta=\{\sigma,\beta,\gamma,\kappa\}$, while $\alpha=4.5$ is a fixed scaling ratio between the two exponential widths.
%mitigate the unavoidable degeneracy between the $\epsilon^\beta$ and $\epsilon^{2\,\gamma}$ terms.

\begin{figure*}
\centering
\includegraphics[width=0.9\textwidth]{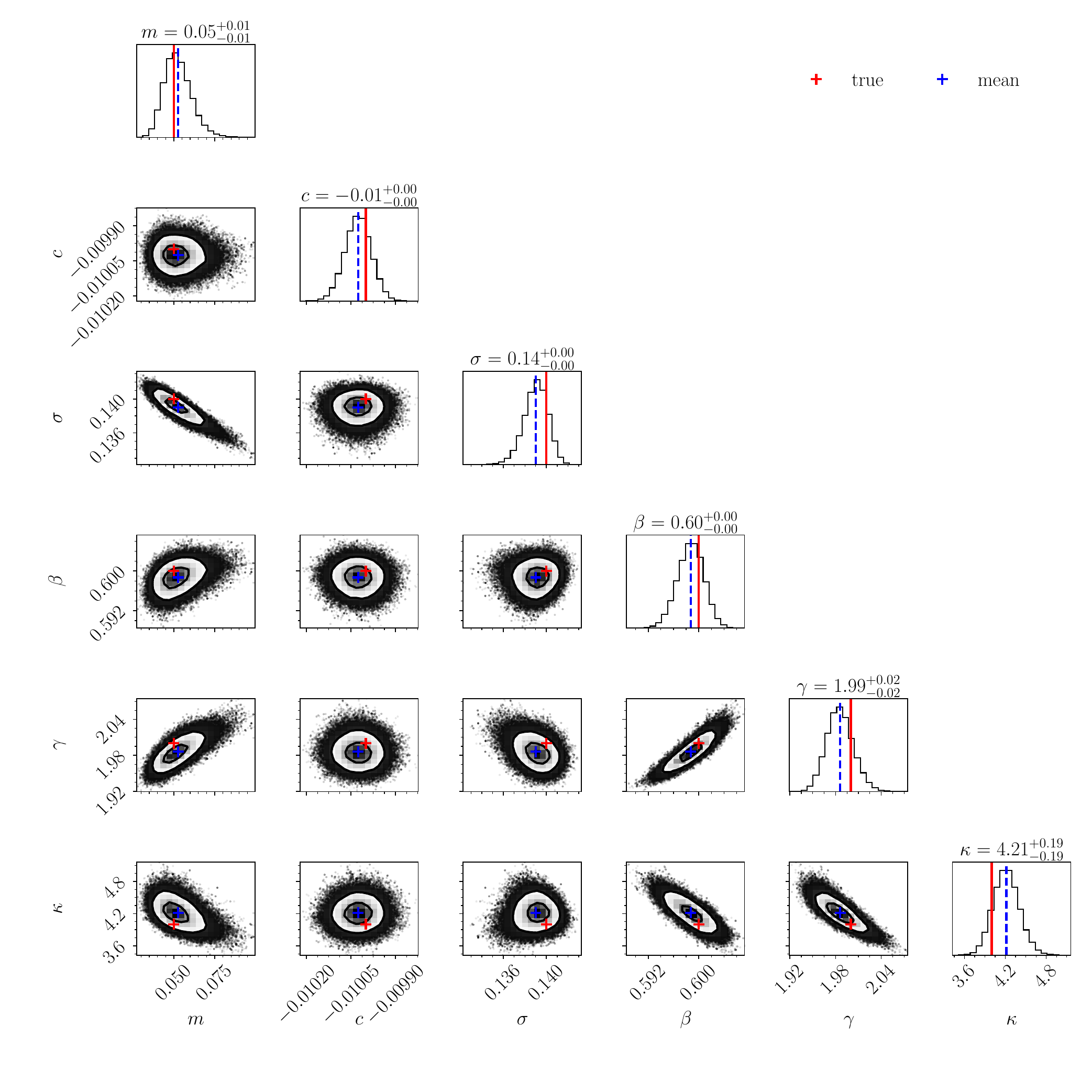}
\vspace{-10pt}
\caption{Joint MCMC samples of the inferred bias parameters ($m$ and $c$) and hyper-parameters of the lensed distribution ($\sigma$, $\beta$, $\gamma$, and $\kappa$).
The red lines/crosses represent the true values, and the blue ones represent the recovered mean values marginalised over nuisance parameters.
Despite the degeneracies, the marginalised $m$ and $c$ are constrained within one standard deviation (in particular, see the two marginal distributions at the top left).}
\label{fig:mcmc}
\end{figure*}

In order to ease the interpretation, we chose some representative, large values for shear biases, $m_\text{true}=0.05$ and $c_\text{true}=-0.01$, and hyper-parameters, $\sigma_\text{true}=0.3$, $\beta_\text{true}=0.6$, $\gamma_\text{true}=2$, and $\kappa_\text{true}=4$ as ground truth, which and based on the Q1 data (see Fig.\;\ref{fig:distributions}).
We simulated $10^7$ ellipticities from the distribution of observed ellipticity [see Eq.\;\eqref{eq:cons_prob}], initially in the absence of noise, and calculated their empirical distribution, $\hat{p}_{\hat{\epsilon}}$, in 1000 bins, as shown in Fig.\;\ref{fig:distributions}.
We sampled the posterior of Eq.\;\eqref{eq:posterior} with the likelihood of Eq.\;\eqref{eq:likelihood} and a wide uniform prior for biases and hyper-parameters: $m\in(-0.1,0.1)$, $c\in(-0.1,0.1)$, $\sigma\in(0.05,0.6)$, $\beta\in(0.1,2)$, $\gamma\in(0.5,4)$, and $\kappa\in(0.5,6)$.
We used the affine-invariant version of the \textsc{LensMC} core sampler \citepalias{congedo2024} and ran 32 parallel chains of 10,000 samples after a short burn-in phase.
Figure\;\ref{fig:mcmc} shows the resulting chains with recovered mean values all within one standard deviation from the ground truth.
Also, there appears to be a degree of correlation between $\sigma$ and $m$ and, to some extent, also between hyper-parameters,
which can be explained in terms of those parameters being all proxies for the dispersion of the distribution.
However, those degeneracies do not hinder the recovery of shear biases, thanks to marginalisation.
Similar conclusions can also be drawn for smaller input biases.
Finally, as a sanity check, we tested the sampling with a Gaussian likelihood (where the variance term is given by the observed distribution) and found fully consistent results, which corroborates the fact that the Poisson distribution converges to a Gaussian very rapidly, the two being practically indistinguishable for a number of samples per bin of about 10.

The uncertainties reported in all parameters reflect the width of the posterior after a wide, flat prior was applied.
All parameters are constrained within one standard deviation and, in particular, $m-m_\text{true}=(2.7\pm8.4)\times10^{-3}$ and $c-c_\text{true}=(-2.6\pm4.4)\times10^{-5}$.
However, a few parameters show some degree of degeneracy (most notably, $m$ and $\sigma$), made gradually worse by a smaller and smaller number of data points.
Since the likelihood scales as $\exp(-N)$, we tested the dependence of the biases on the recovered $m$ and $c$ on the number of samples, $N$, and found that the biases converge to smaller and smaller values, with a residual error on $m$ of less than 1\% for $N>10^6$.

Furthermore, we tested the impact of mismodelling the distribution by purposely ignoring a measurement noise term, $\sigma_{\epsilon_\text{n}}$, in the likelihood, which was actually present in the simulations.
We found that the reduced $\chi^2$ increases quadratically as a function of $\sigma_{\epsilon_\text{n}}$, with a residual error on $m$ of 1\% for $\sigma_{\epsilon_\text{n}}=0.05$.
In fact, as Fig.\;\ref{fig:distributions} shows, the distributions of observed lensed ellipticity and unbiased lensed ellipticity (whose difference accounts for measurement noise and bias) are practically indistinguishable, with a mean $\sigma_{\epsilon_\text{n}}$ of only 0.05, as estimated in the measurement \citepalias{congedo2024}.
This quantity is further reduced to 0.03 on actual real data from Q1 \citep{congedo2026}, which gives us reassurance that, even if this term were completely ignored, the residual bias on $m$ would still be smaller than 1\%.
In general, for the application to real data, the best approach would be to obtain $\sigma_{\epsilon_\text{n}}$ from the measurements and then apply it in the inference of biases.

The potential of this methodology is that selection biases are already correctly accounted for in the observed lensed distribution, which could be defined for a broad sample of galaxies (including tomographic bins, field-of-view positions, or morphological bins).
The natural products of the inference are the effective $m$ and $c$ biases after measurement and selection effects, and marginalised over the hyper-parameters of the unbiased lensed distribution. 
However, the main limitation of the methodology is the accurate modelling of the unbiased lensed distribution.
With real data, the model for the unbiased lensed distribution is not completely known beforehand.
However, the ellipticity has to satisfy specific symmetries and boundary conditions, helping to constrain the functional form of the model.
Therefore, in general, the family of functional forms should be broadly known, and we showed that the same functional form can in fact reproduce both pre-launch simulations and real data with good fidelity (see Fig.\;\ref{fig:distributions}).
This suggests that a sufficiently accurate functional form can always be found for real data.

Moreover, the multiplicative and additive biases of a well-behaved shear measurement method should usually be very small, which is already the case for Stage-III and, to a greater extent, also for Stage-IV surveys.
In the limit of small, percent-level biases, the effect of biases on the lensed ellipticity distribution is perturbative, which implies that the observed distribution will not change significantly from the unbiased distribution (see Fig.\;\ref{fig:distributions}).
Additionally, we observed that the marginalisation mitigates the degeneracy between biases and the hyper-parameters of the distribution (see Fig.\;\ref{fig:mcmc}).
All of this gives us reassurance that, under specific assumptions, it should always be possible to constrain the model from the data already available, and constrain the bias successfully.
Deep fields could also be used to help constrain the functional form of the unbiased lensed distribution as a function of bins of interest even more precisely, provided that selections were applied similarly to the wide fields to account for selection biases properly.
However, the analysis of deep fields and knowledge of deeper distributions are not, by all means, a prerequisite of our method but would only provide a potential aid. 

\section{Conclusions}\label{sec:conclusions}

In the context of cosmic shear surveys, we proposed a novel methodology to infer both multiplicative and additive shear biases directly from the observed lensed distribution of galaxy ellipticities.
Although the estimation of additive biases is routinely applied to real data and there is a long history of measuring shear responses on data, in this paper we provided solid theoretical grounds for the initial ansatz that multiplicative biases could also be inferred empirically from observed distributions alongside additive biases.
Using complementary statistical techniques, i.e., the transformation of probability distribution functions, their moment-generating function, and their raw and central moments, we showed that multiplicative and additive biases impact not only the mean and variance of the distribution, as has been commonly assumed, but all higher moments to any order.
Empirical distributions of galaxy ellipticities do help in that regard because they show a peak at zero ellipticity (potentially with a cusp), a rapid decay for increasing ellipticity,
and are also truncated at $|\epsilon|<1$ (which is a requirement of $\epsilon$-ellipticities transforming under shear, leaving the bounds unchanged).
Those distributions are highly non-Gaussian and therefore contain a lot more information than Gaussian distributions do.
Therefore, the information on the biases is fully encoded in the empirical distribution to any orders and is ready to be exploited.

We introduced statistical modelling of the biased distribution, the model of observed lensed distribution (flexible enough to reproduce \textit{Euclid} pre-launch simulations and real data), the corresponding likelihood, and a Bayesian framework to infer biases and hyper-parameters of the measured lensed distribution jointly.
After simulating representative data with the injection of biases and noise, we found, not so surprisingly, that additive and multiplicative biases can be constrained with enough statistical power that the inferred joint distributions are not strongly degenerate with the hyper-parameters.
Meanwhile, some of the hyper-parameters are indeed correlated.
However, we observed that the marginalisation significantly mitigates the sensitivity of the inferred biases on those degeneracies between hyper-parameters.
Additionally, we studied the dependence of biases on inferred biases on the number of samples and Gaussian noise not captured in the modelling, which could potentially hinder the inference by introducing large model biases.
We demonstrated that the biases could still be recovered with enough statistical power under reasonable conditions.
Finally, we studied the potential limitations, and ways to address them with real data.
This includes the knowledge of the unbiased lensed distribution, which we argue could be constrained from data, and then biases could be fitted jointly with the hyper-parameters.
We demonstrated how the same functional form can reproduce \textit{Euclid} pre-launch simulations and real data, and we showed how the noise term contributing to the observed distribution is fairly small in either case.

We envisage that our methodology is very general and has a wide range of applicability to weak lensing surveys.
Therefore, it warrants further investigation on real data from Stage-IV surveys, especially \textit{Euclid} and \textit{Rubin} where the accuracy requirements are stringent and measurements have to be calibrated with simulations.
Being agnostic of cosmology and galaxy morphology, our self-calibration of cosmic shear biases could not only aid the calibration with simulations by providing useful empirical priors for biases, but also alleviate the reliance on large-scale, realistic simulations, which are usually one of the most difficult tasks of lensing analyses.

\section*{Acknowledgements}

GC and ANT acknowledge support provided by the United Kingdom Space Agency.
ANT acknowledges support provided by the Science and Technology Facilities Council.
Figure\;\ref{fig:mcmc} was made with \texttt{corner} \citep{foreman-mackey2016}.

\section*{Data Availability}

The data was generated with original code, available on request.

\bibliographystyle{mnras}
\bibliography{references.bib}

% Don't change these lines
\bsp	% typesetting comment
\label{lastpage}
\end{document}